\documentclass[aps,showpacs]{revtex4}
\begin{document}
\title{Some exact results on CP and CPT violations in a  $C=-1$ entangled   pseudoscalar neutral meson pair}
\author{Yu Shi}
\email{yushi@fudan.edu.cn}
\affiliation{Center for Field Theory and Particle Physics, Department of Physics, Fudan University, Shanghai
200433, China}

\begin{abstract}
We consider neutral pseudoscalar mesons in an entangled or Einstein-Podolsky-Rosen state with  $C=-1$. Due to quantum entanglement and antisymmetry of  this state, the rates of the joint decays of the meson pair display various interesting features, as is well known. As functions of CP and CPT violating parameters, here we obtain some exact results about the joint decay rates and their asymmetries for a given time difference defined for joint decays to flavor eigenstates, as well as those for joint decays to CP eigenstates. The entanglement allows a meaningful and useful definition of the transition amplitude from a CP eigenstate of a meson.  These results yield useful information and criteria  on CP and CPT violations.
\end{abstract}


European Physical Journal C {\bf 73}, 2506 (2013).

\maketitle

\section{Introduction}

Pseudoscalar neutral mesons are ideal systems in studying CP violation, and in search for CPT violation, which is implied by the standard model extension~\cite{kos1}.
Pseudoscalar neutral mesons in an entangled state have special properties due to quantum entanglement or Einstein-Podolsky-Rosen correlation~\cite{epr,old,test,cern,go,shi1}. Nowadays, such entangled meson pairs  are routinely produced  in $\phi$ or $B$ factories~\cite{domenico,kloe,babar,belle}.  Hence it is interesting and important to explore the use of the entangled pairs in  CP and CPT problems~\cite{domenico,kloe,babar,belle,bernabeu1,dunietz,
buchanan,dambrosio,kobayashi,kos2,bernabeu2,baji,bernabeu,petrov,bigi,
soni,kittle,huang,shi2}.

Previous investigations often used some  approximations up to the first order of the CP or CPT violating parameters. Reasonable as it is, it is of special value to have some exact results without approximations, since  being exactly zero is qualitatively different from being approximately zero with higher orders of CP or CPT violating parameters neglected. Moreover, exact results are important in making comparisons between different approximations.

In this paper, we present some exact results on the use of  the $C=-1$ entangled state of pseudoscalar mesons to examine CP and CPT violations or conservations. A recent calculation ignored direct CP violation and the violation of $\Delta {\cal F} =\Delta Q$ rule, where ${\cal F}$ is the flavor quantum number~\cite{shi2}.

The rest of this paper is organized as the following. In Sec.~\ref{review}, we review the single-particle bases,  the time evolution of a single neutral meson as well as that of an  entangled neutral meson pair with total $C=-1$.  In Sec.~\ref{flavor}, we consider the decays   of both mesons into flavor eigenstates, for which some results concerning  CP and CPT  are given in  Sec.~\ref{violation}.  In Sec.~\ref{cp},  we consider decays   of both mesons into CP eigenstates, for which some results  concerning  CP and  CPT are given  in   Sec~\ref{cp2}.  A summary is made in Sec.~\ref{summary}.

\section{A Review of Single Neutral Meson and Entangled Neutral Mesons   \label{review}}

A neutral pseudoscalar meson $M^0$ and its antiparticle $\bar{M}^0$ are  eigenstates
of parity $P$ both with eigenvalue $-1$, and of a characteristic flavor  ${\cal F}$ with eigenvalues $\pm 1$.   ${\cal F}$  is strangeness for $K^0$ and $\bar{K}^0$, beauty for
$B_d^0$ and $\bar{B}_d^0$, charm for $D^0$ and $\bar{D}^0$, and strangeness or beauty (with a minus sign) for   $B_s^0$ and $\bar{B}_s^0$. With the phase convention
$C|M^0\rangle = -|\bar{M}^0\rangle$ and
$C|\bar{M}^0\rangle = -|M^0\rangle$,   the eigenstates of $CP$ are
\begin{equation}
|M_\pm \rangle =
\frac{1}{\sqrt{2}}(|M^0\rangle \pm  |\bar{M}^0\rangle), \label{cptof}
\end{equation}
with eigenvalues $\pm 1$. It should be noted that physically a single particle cannot be in the state $|M_\pm \rangle$, as CP is violated.

In the flavor basis, the effective mass matrix $H$ can be written as
\begin{equation}
H=\left(
\begin{array}{cc}
H_{00} & H_{0\bar{0}} \\
H_{\bar{0}0} & H_{\bar{0}\bar{0}}
\end{array}\right),
\end{equation}
where $H_{00} \equiv \langle M^0|H|M^0\rangle$,  $H_{0\bar{0}}\equiv\langle M^0|H|\bar{M}^0\rangle $,  $H_{\bar{0}0 } \equiv \langle\bar{ M}^0|H|M^0\rangle$, $H_{\bar{0}\bar{0}} \equiv \langle\bar{ M}^0|H|\bar{M}^0\rangle$.

Indirect T violation and  CP violation are  characterized by a nonzero parameter $\epsilon_M$ defined through
\begin{equation}
\frac{q}{p} \equiv \sqrt{ \frac{H_{\bar{0}0}}{ H_{0\bar{0}}} } \equiv \frac{1-\epsilon_M}{1+\epsilon_M},
\end{equation}
because if CP or T is conserved indirectly, then $\epsilon_M =0$.

Indirect CPT violation and CP violation  are characterized by a nonzero parameter $\delta_M$ defined as~\cite{chou}
\begin{equation}
\delta_M \equiv \frac{ H_{\bar{0}\bar{0}} -H_{00}}{ \sqrt{ H_{0\bar{0}}H_{\bar{0}0} } } \neq 0,
\end{equation}
because if CPT or CP is conserved indirectly, then $\delta_M =0$.

The eigenvalues of $H$ are
\begin{eqnarray}
\lambda_S  \equiv m_S -i\Gamma_S/2= H_{00} +  \sqrt{ H_{0\bar{0}}H_{\bar{0}0} } (\sqrt{1+\frac{\delta_M^2}{4}} + \frac{\delta_M}{2}), \\
\lambda_L \equiv  m_L -i\Gamma_L/2
= H_{\bar{0}\bar{0}}-\sqrt{ H_{0\bar{0}}H_{\bar{0}0} }
(\sqrt{1+\frac{\delta_M^2}{4}} + \frac{\delta_M}{2}),
\end{eqnarray}
corresponding, respectively, to the eigenstates
\begin{eqnarray}
|M_S\rangle =
\frac{1}{\sqrt{|p_S|^2+|q_S|^2}}(p_S|M^0\rangle
+q_S|\bar{M}^0\rangle) = \frac{1}{\sqrt{1+|\epsilon_S|^2}}(|M_+\rangle
+\epsilon_S|M_-\rangle), \label{ms1}\\
|M_L\rangle  =
\frac{1}{\sqrt{|p_L|^2+|q_L|^2}}(p_L|M^0\rangle - q_L|\bar{M}^0\rangle) =  \frac{1}{\sqrt{1+|\epsilon_L|^2}}(
\epsilon_L|M_+\rangle+ |M_-\rangle), \label{ml1}
\end{eqnarray}
with
\begin{eqnarray}
x_S \equiv \frac{q_S}{p_S} \equiv \frac{1-\epsilon_S}{1+\epsilon_S}=
\frac{q}{p}(\sqrt{1+\frac{\delta_M^2}{4}}+\frac{\delta_M}{2}),\\
x_L \equiv \frac{q_L}{p_L} \equiv \frac{1-\epsilon_L}{1+\epsilon_L}=
\frac{q}{p}(\sqrt{1+\frac{\delta_M^2}{4}}-\frac{\delta_M}{2}),
\end{eqnarray}
If $\epsilon_S=0$, $|M_S\rangle=|M_+\rangle$, with $CP |M_+\rangle = |M_+\rangle$. If $\epsilon_L=0$,  $|M_L\rangle = |M_-\rangle$, with  $CP |M_-\rangle = -|M_-\rangle$. Conversely
\begin{eqnarray}
|M_+\rangle =
\frac{1}{1-\epsilon_S\epsilon_L}(\sqrt{1+|\epsilon_S|^2}|M_S\rangle-
\epsilon_S\sqrt{1+|\epsilon_L|^2}|M_L\rangle), \\
|M_-\rangle =
\frac{1}{1-\epsilon_S\epsilon_L}(\sqrt{1+|\epsilon_L|^2}|M_L\rangle-
\epsilon_L\sqrt{1+|\epsilon_S|^2}|M_S\rangle). \label{lm2}
\end{eqnarray}

We often have the definitions
\begin{equation}
\epsilon \equiv \frac{1}{2}(\epsilon_S+\epsilon_L),
\end{equation}
\begin{equation}
\delta \equiv \frac{1}{2}(\epsilon_S-\epsilon_L).
\end{equation}
Hence
\begin{equation}
\epsilon = \frac{\epsilon_M}{1+\epsilon_M^2+(1-\epsilon_M^2)\sqrt{1+\frac{\delta_M^2}{4}}}
\approx \frac{\epsilon_M}{2},
\end{equation}
\begin{equation}
\delta   = -\frac{(1-\epsilon_M^2)\delta_M}{1+\epsilon_M^2+(1-\epsilon_M^2)
\sqrt{1+\frac{\delta_M^2}{4}}}\approx -\frac{\delta_M}{2}.
\end{equation}

Under the Wigner-Weisskopf approximation,
the evolution of an arbitrary state of a pseudoscalar meson $|M(t)\rangle$, as a superposition of $|M^0\rangle$ and $|\bar{M}^0\rangle$,  can be described by a Schr\"{o}dinger equation
\begin{equation}
i\frac{\partial}{\partial t}|M(t)\rangle = H |M(t)\rangle,
\end{equation}
using which we can find the following time-dependent states.

Starting as the mass eigenstate $|M_S\rangle$, the state  of a single meson  evolves as
\begin{equation}
|M_S(t) \rangle = e^{-i\lambda_S t} |M_S\rangle.
\end{equation}
Starting as the mass eigenstate $|M_L\rangle$, the state evolves as
\begin{equation}
|M_L(t) \rangle  =  e^{-i\lambda_L t} |M_L\rangle.
\end{equation}

Starting as the flavor eigenstate $|M^0\rangle$, the state evolves as
\begin{equation}
|M^0(t) \rangle = G_{00}(t) |M^0\rangle
+G_{0\bar{0}}(t)|\bar{M}^0\rangle, \label{m0t}
\end{equation}
where
\begin{eqnarray}
G_{00}(t) &\equiv& \frac{e^{-i\lambda_S t}+\Omega e^{-i\lambda_L t}}{1+\Omega},\\ G_{0\bar{0}}(t)& \equiv &\frac{x_S(e^{-i\lambda_S t}- e^{-i\lambda_L t})}{1+\Omega},
\end{eqnarray}
with
\begin{equation}
\Omega \equiv \frac{x_S}{x_L} = \frac{q_Sp_L}{p_Sq_L}.
\end{equation}
Starting as the flavor eigenstate  $|\bar{M}_0\rangle$, the state evolves as
\begin{equation}
|\bar{M}^0(t) \rangle=
G_{\bar{0}0}(t) |M^0\rangle
+G_{\bar{0}\bar{0}}(t)|\bar{M}^0\rangle, \label{m0bart}
\end{equation}
where
\begin{eqnarray}
G_{\bar{0}0}(t)& \equiv &\frac{x_S^{-1}(e^{-i\lambda_S t}- e^{-i\lambda_L t})}{1+\Omega^{-1}},\\
G_{\bar{0}\bar{0}}(t)& \equiv&
\frac{ e^{-i\lambda_S t}+ \Omega^{-1} e^{-i\lambda_L t}}{1+\Omega^{-1}}.
\end{eqnarray}

Physically an isolated single meson state cannot start as a CP eigenstate $|M_{\pm}\rangle$, as CP is indeed violated. There is no way of tagging one meson to be in such an eigenstate by measuring the other if the two are prepared as  an entangled pair.   However, as explained below, the initial state of a pair of entangled mesons with $C=-1$ is exactly a superposition of two terms, in each of which the two mesons are in different CP eigenstates. Therefore it is useful to consider the evolution of the state of a meson  starting as a CP eigenstate.
Starting as the CP eigenstate $|M_{+}\rangle$, the state evolves as
\begin{equation}
|M_{+}(t)\rangle = F_{+ +}(t) |M_+\rangle + F_{ +-}(t) |M_-\rangle. \label{mtime}
\end{equation}
where
\begin{eqnarray}
F_{++}( t) &=& \frac{1}{2(1+\Omega)}[(1 +x_L^{-1} + x_S + \Omega) e^{-i\lambda_S  t} +(1 -x_L^{-1} - x_S + \Omega) e^{-i \lambda_L  t}],\\
F_{+-}(t) & =& \frac{1 + x_L^{-1} - x_S - \Omega}{2(1+\Omega)}(e^{-i\lambda_S t}- e^{-i \lambda_L t}).
\end{eqnarray}
Starting as the CP eigenstate $|M_{-}\rangle$, the state evolves as
\begin{equation}
|M_{-}(t)\rangle = F_{- +}(t) |M_+\rangle + F_{- -}(t) |M_-\rangle. \label{mtime2}
\end{equation}
where
\begin{eqnarray}
F_{-+}( t)& =& \frac{1 - x_L^{-1} + x_S - \Omega}{2(1+\Omega)}(e^{-i\lambda_S t} - e^{-i \lambda_L  t}),\\
F_{--}( t) &=& \frac{1}{2(1+\Omega)}[(1 - x_L^{-1}- x_S + \Omega) e^{-i\lambda_S  t} +(1 + x_L^{-1} + x_S + \Omega) e^{-i \lambda_L  t}].
\end{eqnarray}

Now consider the $C=-1$ entangled state of a pair of pseudoscalar mesons, which may be  referred to as  Alice (a) and Bob (b),
\begin{eqnarray}
|\Psi_-\rangle & = & \frac{1}{\sqrt{2}}(|M^0\rangle_a|\bar{M}^0\rangle_b
-|\bar{M}^0\rangle_a|M^0\rangle_b),
\end{eqnarray}
which can be produced from a source of $J^{PC}=1^{--}$.

A remarkable feature of $|\Psi_-\rangle$ is that it is also  {\em exactly}
\begin{eqnarray}
|\Psi_-\rangle &=& \frac{1}{\sqrt{2}}(|M_-\rangle_a|M_+\rangle_b-
|M_+\rangle_a|M_-\rangle_b). \label{pm}
\end{eqnarray}
This means that although a single meson cannot be in the state $|M_{\pm}\rangle$, an entangled pair of mesons with $C=-1$ is {\em exactly} in the state  $\frac{1}{\sqrt{2}}(|M_-\rangle_a|M_+\rangle_b-
|M_+\rangle_a|M_-\rangle_b)$. There is no tagging  here for the initial mesons, as only the final decay products are measured. This allows the use of (\ref{mtime}) and (\ref{mtime2})  for the entangled pair, because of the linearity of quantum evolution and that the two entangled but separated mesons are non-interacting.

Starting as $|\Psi_-\rangle$, the state of the entangled meson pair evolves and decays to or produces   certain products at $t_a$ and $t_b$,  which may or may not be equal.  To account for this situation, one has
\begin{eqnarray}
|\Psi_-(t_a,t_b)\rangle &=&\frac{1}{\sqrt{2}}(|M^0(t_a)\rangle_a|\bar{M}^0(t_b)\rangle_b
-|\bar{M}^0(t_a)\rangle_a|M^0(t_b)\rangle_b)\\
&=&\frac{1}{\sqrt{2}}(|M_-(t_a)\rangle_a|M_+(t_b)\rangle_b
-|M_+(t_a)\rangle_a|M_-(t_b)\rangle_b).
\end{eqnarray}

The joint  rate that Alice decays to $|\psi_a\rangle$ at $t_a$ while Bob decays to $|\psi_b\rangle$ at $t_b$ is
\begin{equation}
I(\psi_a,t_a;\psi_b,t_b) = |\langle \psi_a,\psi_b|{\cal H}_a {\cal H}_b |\Psi_-(t_a,t_b)\rangle|^2,
\end{equation}
where ${\cal H}_a$ and ${\cal H}_b$ represent the Hamiltonians governing the decays of $a$ and $b$, respectively,
\begin{eqnarray}
\langle\psi_a,\psi_b|{\cal H}_a {\cal H}_b|\Psi_-(t_a,t_b)\rangle
&=&\frac{1}{\sqrt{2}}(\langle\psi_a|{\cal H}_a|M^0(t_a)\rangle_a \langle \psi_b|{\cal H}_b |\bar{M}_0(t_b)\rangle_b
- \langle\psi_a|{\cal H}_a|\bar{M}_0(t_a)\rangle_a \langle \psi_b |{\cal H}_b |M^0(t_b)\rangle_b) \\
&=&\frac{1}{\sqrt{2}}(\langle\psi_a|{\cal H}_a|M_-(t_a)\rangle_a \langle \psi_b|{\cal H}_b |M_+(t_b)\rangle_b
- \langle\psi_a|{\cal H}_a|M_+(t_a)\rangle_a \langle \psi_b |{\cal H}_b |M_-(t_b)\rangle_b).
\end{eqnarray}

\section{Decays into flavor eigenstates \label{flavor} }

Suppose from the entangled state $|\Psi_-\rangle$, Alice and Bob each decays or transits to a two-valued flavor eigenstate.
Let us generically denote the flavor eigenstates as $|l^{\pm}\rangle$, with eigenvalue $\pm 1$, respectively.  Examples for  $|l^+\rangle$ include the semileptonic decay products $M^-\bar{l}\nu$,  $D^-D_S^+$, $D^-K^+$, $\pi^-D_S^+$, $\pi^-K^+$ from $M^0=B^0$, and  $D_S^-\pi^+$,   $D_S^-D^+$,  $K^-\pi^+$,  $K^-D^+$ from $M^0=B_S^0$.   Examples for  $|l^-\rangle$ include the semileptonic decay products $M^+ l\bar{\nu}$, $D^+D_S^-$, $D^+K^-$, $\pi^+D_S^-$, $\pi^+K^-$ from $\bar{M}^0=B^0$, and  $D_S^+\pi^-$,   $D_S^+D^-$,  $K^+\pi^-$,  $K^+D^-$ from $M^0=B_S^0$. In the CPLEAR experiment on kaons~\cite{cern}, $|l^+\rangle$ and $|l^-\rangle$ are  products produced via interaction with bound nucleons.

One can calculate the amplitude of such a joint decay in which Alice  decays to $| l_a^x\rangle $  at $t_a$ while Bob decays to $|l_b^y\rangle $  at $t_b$,  where $x$ and $y$ each represents $\pm 1$,
\begin{eqnarray}
\langle l_a^x,l_b^y|{\cal H}_a{\cal H}_b|\Psi_-(t_a,t_b)\rangle & = &   \frac{1}{\sqrt{2}}(\langle l_a^x|{\cal H}_a|M^0(t_a)\rangle_a \langle l_b^y|{\cal H}_b|\bar{M}^0(t_b)\rangle_b
-\langle l_a^x|{\cal H}_a|\bar{M}^0(t_a)\rangle_a  \langle l_b^y|{\cal H}_b|M^0(t_b)\rangle_b) \\
&=&  C(l_a^x,l_b^y) e^{-i(\lambda_St_a+\lambda_Lt_b)} + D(l_a^x,l_b^y) e^{-i(\lambda_Lt_a+\lambda_St_b)},
\end{eqnarray}
with
\begin{eqnarray}
 C(l_a^x,l_b^y)  & \equiv & \frac{1}{\sqrt{2}(1+\Omega)}(-x_L^{-1}r_a^x r_b^y + r_a^x\bar{r}_b^y - \Omega \bar{r}_a^x r_b^y + x_S \bar{r}_a^x\bar{r}_b^y), \\
 D(l_a^x,l_b^y)  & \equiv & \frac{1}{\sqrt{2}(1+\Omega)}(x_L^{-1}r_a^x r_b^y +\Omega  r_a^x\bar{r}_b^y - \bar{r}_a^x r_b^y - x_S \bar{r}_a^x\bar{r}_b^y),
\end{eqnarray}
where
\begin{eqnarray}
r^{x}_\alpha & \equiv & \langle l^x_\alpha|{\cal H}_\alpha|M^0\rangle_\alpha,\\
\bar{r}^{x}_\alpha & \equiv &  \langle l^x_\alpha|{\cal H}_\alpha|\bar{M}_0\rangle_\alpha,
\end{eqnarray}
$\alpha=a,b$.

Therefore we obtain the joint rate
\begin{equation}
\begin{array}{rcl}
I(l^x_a, t_a; l^y_b, t_b)&=&   e^{-(\Gamma_S+\Gamma_L)t_a}
\{|C|^2 e^{-\Gamma_L\Delta t}+ |D|^2 e^{-\Gamma_S\Delta t} \\
&& + 2e^{-\frac{\Gamma_S+\Gamma_L}{2}\Delta t} [\Re (C^*D)\cos (\Delta m \Delta t) + \Im (C^*D) \sin  (\Delta m \Delta t) ]\}, \\
\end{array}
\end{equation}
where $C\equiv C(l_a^x,l_b^y)$,  $D\equiv D(l_a^x,l_b^y)$.

In experiments, it is more convenient to use the integrated rate, herein defined as
\begin{equation}
I'(l_a^x, l_b^y, \Delta t) = \int_0^\infty I(l_a^x, t_a; l_b^y, t_a+\Delta t) dt_a,
\end{equation}
which is simply given by $I(l^x_a, t_a; l^y_b, t_a+\Delta t)$ as in (\ref{il}),  with $e^{-(\Gamma_S+\Gamma_L)t_a}$ replaced as $1/(\Gamma_S+\Gamma_L)$.

We focus on the situation that  $|l_a^+\rangle = |l_b^+\rangle=|l^+\rangle$ and  $|l_a^-\rangle = |l_b^-\rangle=|l^-\rangle$,  hence
\begin{eqnarray}
r_a^+=r_b^+  & \equiv  & R^+  =  a+b, \\
\bar{r}_a^+=\bar{r}_b^+ &  \equiv& S^+ =  c^* - d^*, \\
r_a^-=r_b^- & \equiv & S^- = c + d, \\
\bar{r}_a^- = \bar{r}_b^-  &\equiv&   R^- = a^*-b^*,
\end{eqnarray}
where the quantities $a$, $b$, $c$ and $d$ are the ones usually defined  in literature~\cite{maiani,domenico}.

If CP is conserved directly, then we have $R^+=R^-$ and $S^+=S^-$. If CPT is conserved directly, then we have $(R^+)^*=R^-$ and $(S^+)^*=S^-$. If $\Delta {\cal F} = \Delta Q$ rule is respected, then we have $S^\pm =0$.

One obtains
\begin{eqnarray}
C(l^+,l^+)  & = & -D(l^+,l^+)= \frac{1}{\sqrt{2}(1+\Omega)}
[-x_L^{-1}(R^+)^2 +(1-\Omega) R^+S^+ +x_S (S^+)^2], \\
C(l^-,l^-)  & = & -D(l^+,l^+)= \frac{1}{\sqrt{2}(1+\Omega)}
[-x_L^{-1}(S^-)^2 +(1-\Omega) R^-S^- +x_S (R^-)^2], \\
C(l^+,l^-)  & = &  -D(l^-,l^+)  = U  \\
D(l^+,l^-)  & = &  -C(l^-,l^+) =  V,
\end{eqnarray}
where
\begin{eqnarray}
U \equiv   \frac{1}{\sqrt{2}(1+\Omega)}
[-x_L^{-1}R^+ S^- + x_S S^+ R^- + R^+R^- -\Omega S^+ S^-],  \label{u0} \\
V \equiv  \frac{1}{\sqrt{2}(1+\Omega)}
[x_L^{-1}R^+ S^- - x_S S^+ R^- + \Omega R^+R^- - S^+ S^-]. \label{v0}
\end{eqnarray}

Therefore we obtain
\begin{equation}
\begin{array}{rcl}
I(l^+, t_a; l^+, t_b)&=& \frac{|-x_L^{-1}(R^+)^2 +(1-\Omega) R^+S^+ +x_S (S^+)^2|^2}{2|1+\Omega|^2} e^{-(\Gamma_S+\Gamma_L)t_a}
[e^{-\Gamma_L\Delta t} + e^{-\Gamma_S\Delta t} -2e^{-\frac{\Gamma_S+\Gamma_L}{2}\Delta t} \cos (\Delta m \Delta t)], \\
I(l^-, t_a; l^-, t_b)&=& \frac{|-x_L^{-1}(S^-)^2 +(1-\Omega) R^-S^- +x_S (R^-)^2|^2}{2|1+\Omega|^2} e^{-(\Gamma_S+\Gamma_L)t_a}
[e^{-\Gamma_L\Delta t} + e^{-\Gamma_S\Delta t} -2e^{-\frac{\Gamma_S+\Gamma_L}{2}\Delta t} \cos (\Delta m \Delta t)], \\
I(l^+, t_a; l^-, t_b) &=&   e^{-(\Gamma_S+\Gamma_L)t_a}
\{|U|^2 e^{-\Gamma_L\Delta t}+ |V|^2 e^{-\Gamma_S\Delta t} + 2e^{-\frac{\Gamma_S+\Gamma_L}{2}\Delta t} [\Re (U^*V)\cos (\Delta m \Delta t) + \Im (U^*V) \sin  (\Delta m \Delta t) ]\},
\\
I(l^-, t_a; l^+, t_b) &=&   e^{-(\Gamma_S+\Gamma_L)t_a}
\{|V|^2 e^{-\Gamma_L\Delta t}+ |U|^2 e^{-\Gamma_S\Delta t} + 2e^{-\frac{\Gamma_S+\Gamma_L}{2}\Delta t} [\Re (U^*V)\cos (\Delta m \Delta t) - \Im (U^*V) \sin  (\Delta m \Delta t) ]\},
\end{array} \label{il}
\end{equation}
where $\Delta m \equiv m_L-m_S$, $\Delta t \equiv t_b-t_a $.

\section{  CP and CPT violations in joint decays into flavor eigenstates \label{violation}}

The above four joint rates can form some asymmetries among them. As
$
I'(l^x, l^y, \Delta t)e^{-(\Gamma_L+\Gamma_S)t_a} = \frac{1}{\Gamma_L+\Gamma_S} I(l^x, t_a; l^y, t_b),$
the asymmetries defined for the instantaneous joint rate $I(l^x, t_a; l^y, t_a+\Delta t )$
and its integration $I'(l^x, l^y, \Delta t)$ for a specific $\Delta t$ are equal, and depend  only  on $\Delta t $,
\begin{equation}
A(l^{x} l^{y}, l^{z}l^{w}, \Delta t) \equiv \frac{I[l^x,t_a; l^y, t_a+\Delta t ]-I[l^z,t_a; l^w, t_a+\Delta t ]}{I[l^x,t_a; l^y, t_a+\Delta t ]+ I[l^z,t_a; l^w, t_a+\Delta t ]}=  \frac{I'[l^x, l^y, \Delta t]-I'[l^z, l^w, \Delta t]}
{I'[l^x, l^y, \Delta t]+I'[l^z, l^w, \Delta t]}.
\end{equation}

\subsection{Equal-flavor asymmetry }

Now we consider the equal-flavor asymmetry, which is
\begin{equation}
A(++,--, \Delta t) = \frac{|x_S (S^+)^2-x_L^{-1}(R^+)^2 +(1-\Omega) R^+S^+ |^2 - |x_S (R^-)^2-x_L^{-1}(S^-)^2 +(1-\Omega) R^-S^- |^2 }{|x_S (S^+)^2-x_L^{-1}(R^+)^2 +(1-\Omega) R^+S^+ |^2 + |x_S (R^-)^2-x_L^{-1}(S^-)^2 +(1-\Omega) R^-S^- |^2 }, \label{a10}
\end{equation}
which, because of the antisymmetry of $|\Psi_-\rangle$,  is analogous to the famous Kabir asymmetry defined for the difference between the transition rates from $M^0$ to $\bar{M}^0$ and that from $\bar{M}^0$ to $M^0$~\cite{bigi}. Note that $A(++,--, \Delta t)$ is a constant independent of $\Delta t$.

If CP is conserved directly, then $R^+ = R^-$, $S^+ = S^-$, consequently
\begin{equation}
A(++,--, \Delta t) = \frac{|x_S (\zeta_+)^2-x_L^{-1} +(1-\Omega) \zeta_+ |^2 - |x_S-x_L^{-1}(\zeta_+)^2 +(1-\Omega) \zeta_+ |^2 }{|x_S (\zeta_+)^2-x_L^{-1} +(1-\Omega) \zeta_+ |^2 - |x_S-x_L^{-1}(\zeta_+)^2 +(1-\Omega) \zeta_+ |^2  },  \label{cpd}
\end{equation}
where
\begin{equation}
\zeta_\pm \equiv \frac{S^\pm}{R^\pm},
\end{equation}
which characterizes the violation of $\Delta {\cal F} =\Delta Q$ rule.

If CPT is conserved directly, then $(R^+)^*=R^-$ and $(S^+)^*=S^-$, consequently
\begin{equation}
A(++,--, \Delta t) = \frac{|x_S (\zeta_+)^2-x_L^{-1} +(1-\Omega) \zeta_+ |^2 - |x_S^*-(x_L^*)^{-1}(\zeta_+)^2 +(1-\Omega^*) \zeta_+ |^2 }{|x_S (\zeta_+)^2-x_L^{-1} +(1-\Omega) \zeta_+ |^2 + |x_S^*-(x_L^*)^{-1}(\zeta_+)^2 +(1-\Omega^*) \zeta_+ |^2  }.
\end{equation}

We can now obtain $A(++,--, \Delta t)$ for various combination cases.

If both CP and CPT are conserved directly, then $R^+ = R^-$ and $S^+ = S^-$ are both real numbers, consequently  $A(++,--, \Delta t)$ is still given by (\ref{cpd}), now with $\zeta_+$ being real number.

If CP is conserved indirectly, no matter whether CPT is conserved indirectly,  then $\epsilon_M=\delta_M=0$ and thus $x_S=x_L=\Omega=1$,  consequently
\begin{equation}
A(++,--, \Delta t) = \frac{|(S^+)^2-(R^+)^2  |^2 - |(R^-)^2-(S^-)^2 |^2 }{|(S^+)^2-(R^+)^2  |^2 + |(R^-)^2-(S^-)^2 |^2 }.
\end{equation}

If CPT is conserved indirectly, then $ \delta_M=0$, thus $x_S=x_L =q/p$, and thus  $\Omega =1$, consequently
\begin{equation}
A(++,--, \Delta t) = \frac{|\frac{q}{p}(S^+)^2-\frac{p}{q}(R^+)^2  |^2 - |\frac{q}{p}(R^-)^2-\frac{p}{q}(S^-)^2 |^2 }{|\frac{q}{p}(S^+)^2-\frac{p}{q}(R^+)^2  |^2 + |\frac{q}{p}(R^-)^2-\frac{p}{q}(S^-)^2 |^2 }.
\end{equation}

If CP is conserved both directly and indirectly, then
\begin{equation}
A(++,--, \Delta t) = 0. \label{zero1}
\end{equation}

If CPT is conserved both directly and indirectly, then
\begin{eqnarray}
A(++,--, \Delta t) &  = & \frac{|\frac{q}{p}(\zeta_+)^2- \frac{p}{q}   |^2 - |\frac{q^*}{p^*}- \frac{p^*}{q^*}  (\zeta_+)^2   |^2 }{  |\frac{q}{p}(\zeta_+)^2- \frac{p}{q}   |^2 + |\frac{q^*}{p^*}- \frac{p^*}{q^*}  (\zeta_+)^2   |^2 } \\ & = & \frac{(|\frac{q}{p}|^2-|\frac{p}{q}|^2)(|\zeta_+|^4-1)}{
(|\frac{q}{p}|^2+|\frac{p}{q}|^2)(|\zeta_+|^4+1)-4\Re (\frac{qp^*}{pq^*}(\zeta_+)^2)}. \label{a1}
\end{eqnarray}

If CP is conserved directly while CPT is conserved indirectly, then
\begin{eqnarray}
A(++,--, \Delta t) & = &  \frac{|\frac{q}{p}(\zeta_+)^2- \frac{p}{q}   |^2 - |\frac{q}{p}- \frac{p}{q}  (\zeta_+)^2   |^2 }{  |\frac{q}{p}(\zeta_+)^2- \frac{p}{q}   |^2 + |\frac{q}{p}- \frac{p}{q}  (\zeta_+)^2   |^2 }  \\  & = & \frac{(|\frac{q}{p}|^2-|\frac{p}{q}|^2)(|\zeta_+|^4-1)- 4\Im (\frac{qp^*}{pq^*})\Im((\zeta_+)^2) }{
(|\frac{q}{p}|^2+|\frac{p}{q}|^2)(|\zeta_+|^4+1)-4\Re (\frac{qp^*}{pq^*})\Re((\zeta_+)^2)}.
\end{eqnarray}

If CP is conserved indirectly while CPT is conserved directly, then
\begin{equation}
A(++,--, \Delta t) = 0. \label{zero2}
\end{equation}

Combining the above two cases of $A(++,--, \Delta t) = 0$, as stated in (\ref{zero1}) and (\ref{zero2}), we obtain the  following  exact theorems.

{\bf Theorem 1}
   {\em If the equal-flavor asymmetry $A(++,--,\Delta t) \neq 0$,  then  we have one or two of the following violations: (1) CP is violated indirectly, (2) both CP and CPT are violated directly. Therefore assuming direct $CPT$ conservation, nonzero $A(++,--, \Delta t)$ implies that CP must be violated indirectly. }

Moreover, under the assumption that CPT is conserved both directly and indirectly, if  $A(++,--,\Delta t) \neq 0$, then both factors in the numerator in (\ref{a1}) must be nonzero.  Hence we have the following exact theorem, which is clearly consistent with Theorem 1, and is especially useful in testing T violation.

{\bf Theorem 2} {\em If the equal-flavor asymmetry  $A(++,--,\Delta t) \neq 0$ while CPT is assumed to be conserved both directly and indirectly, then in addition to indirect CP violation, we can draw the following conclusions: (1) $|q/p| \neq 1$, i.e.  T must also be  violated indirectly; (2) $|\zeta_+| \neq 1$, i.e.   $|\langle l^+ |{\cal H}|\bar{M}_0\rangle| \neq |\langle l^+|{\cal H}  |M^0\rangle|$,  $|\langle  l^-|{\cal H}  |M^0\rangle | \neq |\langle l^- |{\cal H}|\bar{M}_0\rangle| $, despite   $ \langle l^+|{\cal H}  |M^0\rangle = \langle l^- |{\cal H}|\bar{M}_0\rangle^*$ and $\langle l^+ |{\cal H}|\bar{M}_0\rangle  = \langle  l^-|{\cal H}  |M^0\rangle^* $. }

\subsection{Unequal-flavor asymmetry }

Now we consider the unequal-flavor asymmetry $A(+-,-+, \Delta t)$, which is given in general by
\begin{equation}
A(+-,-+, \Delta t) = \frac{(|U|^2-|V|^2)(e^{-\Gamma_L\Delta t}-e^{-\Gamma_S\Delta t})+4\Im(U^*V)\sin (\Delta m\Delta t)}{(|U|^2+|V|^2)(e^{-\Gamma_L\Delta t}+e^{-\Gamma_S\Delta t})+4\Re(U^*V)\cos (\Delta m\Delta t) }, \label{genuneq}
\end{equation}
with $U$ and $V$ given in (\ref{u0}) and (\ref{v0}).
$A(+-,-+, \Delta t=0)$  always vanishes  exactly.

If CP is conserved directly, then $R^+ = R^-$, $S^+ = S^-$, consequently
\begin{eqnarray}
U & = & \frac{1}{\sqrt{2}(1+\Omega)}
[(x_S-x_L^{-1}) R^+ S^+  + (R^+)^2 -\Omega (S^+)^2], \label{ucp0} \\
V  & = & \frac{1}{\sqrt{2}(1+\Omega)}
[(x_L^{-1}- x_S) R^+ S^+   + \Omega (R^+)^2 - (S^+)^2], \label{vcp0}
\end{eqnarray}
hence $A(+-,-+, \Delta t)$ is given by (\ref{genuneq}) with the following replacement,
\begin{equation}
\begin{array}{rcl}
|U|^2-|V|^2& \rightarrow &  (1-|\Omega|^2)(1-|\zeta_+|^4)+2\Re [(1+\Omega^*)(x_S-x_L^{-1})\zeta_+(1-{\zeta_+^*}^2)] -4 \Im({\zeta_+^*}^2)\Im \Omega, \\
\Im(U^*V) &\rightarrow & \Im[-(x_S-x_L^{-1})\zeta_+(1+\Omega^*)(1-{\zeta_+^*}^2)   -\zeta_+^2 - |\Omega|^2 {\zeta_+^*}^2 +\Omega +  \Omega^*|\zeta_+|^4],\\
|U|^2+|V|^2 &\rightarrow & 2|x_S-x_L^{-1}|^2|\zeta_+|^2  + (1+|\Omega|^2)(1+|\zeta_+|^4)+2\Re [(1-\Omega^*)(x_S-x_L^{-1})\zeta_+(1+{\zeta_+^*}^2)   ] -4 \Re({\zeta_+}^2)\Re \Omega,\\
\Re(U^*V)&\rightarrow & -|x_S-x_L^{-1}|^2|\zeta_+|^2 +\Re[-(x_S-x_L^{-1})\zeta_+(1-\Omega^*)(1+{\zeta_+^*}^2)   -\zeta_+^2 - |\Omega|^2 {\zeta_+^*}^2 +\Omega +  \Omega^*|\zeta_+|^4].
\end{array} \label{sub0}
\end{equation}

Furthermore, if CP is conserved directly while CPT is conserved indirectly, then
\begin{eqnarray}
U & = & \frac{1}{\sqrt{2}(1+\Omega)}
[(\frac{q}{p}-\frac{p}{q}) R^+ S^+  + (R^+)^2 - (S^+)^2], \label{ucp} \\
V  & = & \frac{1}{\sqrt{2}(1+\Omega)}
[(\frac{p}{q}- \frac{q}{p}) R^+ S^+   + (R^+)^2 - (S^+)^2], \label{vcp}
\end{eqnarray}
consequently,
\begin{equation}
A(+-,-+, \Delta t) =
 \frac{2 \Re [(\frac{q}{p}- \frac{p}{q})\zeta_+ (1-{\zeta_+^*}^2)](e^{-\Gamma_L\Delta t}-e^{-\Gamma_S\Delta t})-4\Im[(\frac{q}{p}- \frac{p}{q})\zeta_+(1- {\zeta_+^*}^2)] \sin (\Delta m\Delta t)}{  [ |\frac{q}{p}- \frac{p}{q}|^2 |\zeta_+|^2 +  |1- {\zeta_+}^2|^2 ]  (e^{-\Gamma_L\Delta t}+e^{-\Gamma_S\Delta t})+ 2[ -|\frac{q}{p}- \frac{p}{q}|^2 |\zeta_+|^2 +  |1- {\zeta_+}^2|^2 ] \cos (\Delta m\Delta t) }.
\end{equation}

If CPT is conserved directly, then $(R^+)^*=R^-$ and $(S^+)^*=S^-$, consequently
\begin{eqnarray}
U & = & \frac{1}{\sqrt{2}(1+\Omega)}
[-x_L^{-1} R^+ (S^+)^* +x_S (R^+)^* S^+  + |R^+|^2 -\Omega |S^+|^2], \\
V  & = & \frac{1}{\sqrt{2}(1+\Omega)}
[x_L^{-1} R^+ (S^+)^* - x_S (R^+)^* S^+  + \Omega |R^+|^2 - |S^+|^2],
\end{eqnarray}
hence $A(+-,-+, \Delta t)$ is given by (\ref{genuneq}) with the following replacement, $|U|^2-|V|^2 \rightarrow   (1-|\Omega|^2)(1-|\zeta+|^4)+2\Re\{ x_S  (1+\Omega^*)\zeta_+ - x_L^{-1} (1-\Omega^*)\zeta_+^*  \}(1-|\zeta+|^2)$,   $|U|^2 + |V|^2 \rightarrow  2 (|x_L^{-1}|^2+ |x_S|^2-2 \Re \Omega) |\zeta_+|^2 +  (1+ |\Omega|^2) (1+|\zeta_+|^4) -4 \Re ({x_L^{-1}}^*x_S \zeta_+^2 ) +2 \Re\{ [ (1-\Omega^*)x_S -(1-\Omega) {x_L^{-1}}^*]\zeta_+\} (1 +  |\zeta_+|^2)$,  $U^* V \rightarrow -(1+|\Omega|^2 + |x_S|^2+|x_L^{-1}|^2) |\zeta_+|^2 + \Omega  + \Omega^* |\zeta_+|^4 + {x_L^{-1}}^* x_S {\zeta_+}^2 -(\Omega{x_L^{-1}}^* + x_S) \zeta_+  + ({x_L^{-1}}^*  + \Omega^* x_S ) \zeta_+ |\zeta_+|^2 + (\Omega x_S^* +x_L^{-1} )  {\zeta_+}^*  -( x_S^* +\Omega^* x_L^{-1} )  {\zeta_+}^* |\zeta_+|^2 + x_S^*x_L^{-1}  {{\zeta_+}^*}^2$.

Furthermore, if  CPT is conserved directly while CP is conserved indirectly, then
\begin{eqnarray}
U & = & \frac{1}{2\sqrt{2}}
[-R^+ (S^+)^*  + S^+(R^+)^* + |R^+|^2 - |S^+|^2], \\
V  & = & \frac{1}{\sqrt{2}}
[R^+ (S^+)^* -  S^+(R^+)^*  + |R^+|^2 - |S^+|^2].
\end{eqnarray}
Consequently $A(+-,-+, \Delta t)$ is given by (\ref{genuneq}) under the replacement specified in the last paragraph now with $x_L=x_S=\Omega =1$, therefore
\begin{equation}
A(+-,-+, \Delta t) =
 \frac{ 2(1-|\zeta_+|^2)\Re(\zeta_+)(e^{-\Gamma_L\Delta t}-e^{-\Gamma_S\Delta t})-4(1-|\zeta_+|^2)\Im(\zeta_+) \sin (\Delta m\Delta t)}{   |1-\zeta_+^2|^2 (e^{-\Gamma_L\Delta t}+e^{-\Gamma_S\Delta t})+ 2[ 1+|\zeta_+|^4 -4|\zeta_+|^2 + 2\Re (\zeta_+^2) ] \cos (\Delta m\Delta t) }.
\end{equation}

If both CP and CPT are conserved directly,  $R^+ = R^-$ and $S^+ = S^-$ are both real numbers, consequently $U$ and $V$ are  given as (\ref{ucp0}) and (\ref{vcp0}), while $A(+-,-+, \Delta t)$ is given by (\ref{genuneq})  with the   replacement (\ref{sub0}), but now with  $\zeta_+=\zeta_+^*=\Re(\zeta_+^*)$  while $\Im ({\zeta^*}^2)=0$.

If CPT is conserved indirectly,  then  $x_S=x_L =q/p$,  $\Omega =1$, hence
\begin{eqnarray}
U & = & \frac{1}{2\sqrt{2}}
[-\frac{p}{q} R^+ S^-  + \frac{q}{p} S^+R^- + R^+ R^- - S^+ S^-], \label{upq} \\
V & = & \frac{1}{\sqrt{2}}
[\frac{p}{q}R^+ S^- - \frac{q}{p} S^+R^-  + R^+ R^- - S^+ S^-].  \label{vpq}
\end{eqnarray}
Consequently  $A(+-,-+, \Delta t)$ is given by (\ref{genuneq}) with the following replacement,  $|U|^2-|V|^2 \rightarrow   4 \Re (-\frac{p^*}{q^*}  {\zeta_-}^*  + \frac{p^*}{q^*} |\zeta_-|^2 \zeta_+  + \frac{q^*}{p^*} {\zeta+}^*  -   \frac{q^*}{p^*} |\zeta_+|^2 \zeta_-) $, $|U|^2 + |V|^2 \rightarrow  2 (|\frac{p}{q} \zeta_-|^2 + |\frac{q}{p} \zeta_+ |^2 + 1 + |\zeta_+\zeta_-|^2 ) - 4 \Re ( \frac{p^*q}{q^*p} {\zeta_-}^* \zeta_+  +  \zeta_+ \zeta_-)$, $\Re (U^*V) \rightarrow 1 -|\frac{p}{q} \zeta_-|^2 - |\frac{q}{p} \zeta_+ |^2  + |\zeta_+ \zeta_-|^2 +  2\Re ( \frac{p^*q}{q^*p} \zeta_+{\zeta_-}^*  - \zeta_+\zeta_- ) $,   $\Im (U^*V) \rightarrow    2\Im ( \frac{p}{q} \zeta_- +  \frac{p^*}{q^*} |\zeta_-|^2 \zeta_+ - \frac{q}{p} \zeta_+ - \frac{q^*}{p^*} |\zeta_+|^2\zeta_- ) $.

If CP is conserved indirectly, then no matter whether CPT is conserved indirectly,  $x_S=x_L=\Omega=1$,  consequently $U$ and $V$ are given by (\ref{upq}) and (\ref{vpq}) with $p/q=1$, hence  $A(+-,-+, \Delta t)$ is given by (\ref{genuneq}) with the following replacement,   $|U|^2-|V|^2 \rightarrow   4 \Re (-  {\zeta_-}^*  +  |\zeta_-|^2 \zeta_+  +   {\zeta+}^*  -   |\zeta_+|^2 \zeta_-) $, $|U|^2 + |V|^2 \rightarrow  2 (|  \zeta_-|^2 + | \zeta_+ |^2 + 1 + |\zeta_+\zeta_-|^2) - 8\Re(\zeta_-) \Re(\zeta_+)$, $\Re (U^*V) \rightarrow 1 -| \zeta_-|^2 - | \zeta_+ |^2  + |\zeta_+ \zeta_-|^2 +  4 \Im (   \zeta_+) \Im (\zeta_-)$,   $\Im (U^*V) \rightarrow    2\Im ( \zeta_- +  |\zeta_-|^2 \zeta_+ -   \zeta_+ - |\zeta_+|^2\zeta_- ) $.

If CP is conserved both directly and indirectly,
\begin{equation}
U  = V  =  \frac{1}{2\sqrt{2}}
[(R^+)^2 - (S^+)^2],  \label{uvcp2}
\end{equation}
which, according to (\ref{genuneq}), implies
\begin{equation}
A(+-,-+, \Delta t)=0. \label{cpuneq}
\end{equation}
Hence we have the following exact theorem.

{\bf Theorem 3} {\em If  the unequal-flavor asymmetry
$A(+-,-+,  \Delta t) \neq 0 $, then CP must be violated,  directly or indirectly or both.  }

If CPT is conserved both directly and indirectly, then
\begin{eqnarray}
U & = & \frac{1}{2\sqrt{2} }
[-\frac{p}{q} R^+ (S^+)^* +\frac{q}{p}  (R^+)^* S^+  + |R^+|^2 - |S^+|^2], \\
V  & = & \frac{1}{2\sqrt{2} }
[\frac{p}{q} R^+ (S^+)^* - \frac{q}{p}  (R^+)^* S^+  + |R^+|^2 - |S^+|^2].
\end{eqnarray}
Consequently, in obtaining $ A(+-,-+,  \Delta t)$, the replacement for the four functions of $U$ and $V$  is given by that for the case of direct CPT conservation with $\zeta_-=\zeta_+^*$. Therefore
\begin{equation}
A(+-,-+, \Delta t) = \frac{(1-|\zeta_+|^2)\{ -2\Re [(\frac{p}{q}- \frac{q^*}{p^*})\zeta_-](e^{-\Gamma_L\Delta t}-e^{-\Gamma_S\Delta t})+4\Im[(\frac{p}{q}+ \frac{q^*}{p^*})\zeta_-] \sin (\Delta m\Delta t)}{E_+   (e^{-\Gamma_L\Delta t}+e^{-\Gamma_S\Delta t})+2 E_- \cos (\Delta m\Delta t) },
\end{equation}
where
$E_{\pm}= (1-|\zeta_+|^2)^2 \pm  |\frac{p}{q}\zeta_- - \frac{q}{p}\zeta_+|^2$.
Therefore we have the following exact theorem.

{\bf Theorem 4} {\em If   the unequal-flavor asymmetry
$A(+-,-+,  \Delta t)\neq 0$ for $\Delta t \neq 0$ while CPT is assumed to be conserved both directly and indirectly, then  we can draw the following conclusions: (1)  $|\zeta_+| = |\zeta_-| \neq 1$, i.e.,  $|\langle l^+ |{\cal H}|\bar{M}_0\rangle|= |\langle  l^-|{\cal H}  |M^0\rangle | \neq |\langle l^+|{\cal H}  |M^0\rangle|= |\langle l^- |{\cal H}|\bar{M}_0\rangle| $; (2)  moreover, $\zeta_- =\zeta_+^* \neq 0$, i.e.,  $\langle l^{-}|{\cal H} |M^0\rangle = \langle l^{+}|{\cal H}|\bar{M}^0\rangle^* \neq 0$, which means $\Delta {\cal F} =\Delta Q$ rule must be violated.  }

\section{Decays into CP eigenstates \label{cp} }

Now we consider the situation that the decay products of Alice and Bob  are both CP eigenstates, as in  KLOE experiment, where the rate of both kaons decaying to  $\pi^+\pi^-$ was obtained up to a proportional factor~\cite{kloe,domenico}.  We generically denote the CP eigenstates as $|h^+\rangle$ with eigenvalue $+1$ and  $|h^-\rangle$ with eigenvalue $-1$.  Examples for  $|h^+\rangle$ include    $\pi^+\pi^-$, $\pi^0\pi^0$, etc.    Examples for  $|h^-\rangle$ include  $\pi^0\pi^0\pi^0$, etc. For convenience of discussions, we introduce the parameters defined as
\begin{equation}
w_{\pm\alpha}^x \equiv \langle h^x_{\alpha}|{\cal H}_{\alpha}|M_{\pm}\rangle_\alpha,
\end{equation}
where $x=\pm 1$. These parameters were not introduced previously, because an isolated single meson cannot be in the state $|M_\pm\rangle$ physically. However, the entangled state $|\Psi_-\rangle$ renders  the introduction of these parameters meaningful and convenient, because the entangled state can be exactly written in terms of the CP basis, as in  (\ref{pm}), hence an entangled pair of mesons decay from superposition of direct products of CP eigenstates. No that there is no tagging in CP basis, only the final decay products, e.g. pions,  are measured. Therefore the use of parameter $w_{\pm\alpha}^{x}$ is legitimate.  This is a remarkable point we would like to  exploit. The parameter $w_{\pm}^x$ may be measured by using entangled mesons. They cannot be directly measured by using single mesons, as there is no way to prepare a single meson in a  CP eigenstate. Nevertheless, by using (\ref{cptof}), we have
\begin{equation}
w_{\pm}^x \equiv \frac{1}{\sqrt{2}}(\langle h^x|{\cal H}|M^0\rangle \pm\langle h^x|{\cal H}|\bar{M}^0\rangle),
\end{equation}
where each term on the RHS can be measured for isolated single mesons. Analogously, $w_{\pm}^x$ can also be related to parameters for mass basis,  by using (\ref{lm2}),
\begin{eqnarray}
w_+^x =
\frac{1}{1-\epsilon_S\epsilon_L}(\sqrt{1+|\epsilon_S|^2}\langle h^x|{\cal H}|M_S\rangle-
\epsilon_S\sqrt{1+|\epsilon_L|^2}\langle h^x|{\cal H}|M_L\rangle), \\
w_-^x  =
\frac{1}{1-\epsilon_S\epsilon_L}(\sqrt{1+|\epsilon_L|^2}\langle h^x|{\cal H}|M_L\rangle-
\epsilon_L\sqrt{1+|\epsilon_S|^2}\langle h^x|{\cal H}|M_S\rangle),
\end{eqnarray}
where each term on the RHS of each identity can be measured for isolated single mesons.

One can calculate the  amplitude of such a joint decay in which Alice decays to  $| h_a^x\rangle $  at $t_a$ while Bob decays to $|h_b^y\rangle $  at $t_b$,  where $x$ and $y$ each represents $\pm 1$,
\begin{eqnarray}
\langle h_a^x,h_b^y|{\cal H}_a {\cal H}_b |\Psi_-(t_a,t_b)\rangle & = &   \frac{1}{\sqrt{2}}(\langle h_a^x|{\cal H}_a|M^+(t_a)\rangle_a \langle h_b^y|{\cal H}_b |M_-(t_b)\rangle_b
-\langle h_a^x|{\cal H}_a|M^-(t_a)\rangle_a \langle h_b^y|{\cal H}_b |M_+(t_b)\rangle_b) \\
&=&  M(h_a^x,h_b^y) e^{-i(\lambda_St_a+\lambda_Lt_b)} + N(h_a^x,h_b^y) e^{-i(\lambda_Lt_a+\lambda_St_b)},
\end{eqnarray}
where
$ M(h_a^x,h_b^y)   \equiv  [- (1 -x_L^{-1} + x_S-\Omega) w_{+a}^x w_{+b}^y
+ (1 +x_L^{-1} + x_S+\Omega)w_{+a}^x w_{-b}^y
- (1 -x_L^{-1} - x_S+\Omega)w_{-a}^x w_{+b}^y
+ (1 +x_L^{-1} - x_S-\Omega)w_{-a}^x w_{-b}^y ]/[2\sqrt{2}(1+\Omega)]$, $
N(h_a^x,h_b^y) \equiv  [ (1 -x_L^{-1} + x_S-\Omega)w_{+a}^x w_{+b}^y + (1- x_L^{-1}- x_S+\Omega) w_{+a}^x w_{-b}^y -
(1 +x_L^{-1} + x_S+\Omega)w_{-a}^x w_{+b}^y
- (1 +x_L^{-1} - x_S-\Omega)w_{-a}^x w_{-b}^y ]/[2\sqrt{2}(1+\Omega)]$.

Therefore we obtain the joint rate
\begin{equation}
\begin{array}{rcl}
I(h^x_a, t_a; h^y_b, t_b)&=&   e^{-(\Gamma_S+\Gamma_L)t_a}
\{|M |^2 e^{-\Gamma_L\Delta t}+ |N |^2 e^{-\Gamma_S\Delta t} \\
&& + 2e^{-\frac{\Gamma_S+\Gamma_L}{2}\Delta t} [\Re (M^* N )\cos (\Delta m \Delta t) + \Im (M^*N) \sin  (\Delta m \Delta t) ]\}, \\
\end{array}
\end{equation}
where $M\equiv M(h_a^x,h_b^y)$, $N\equiv N(h_a^x,h_b^y)$.

In experiments, it is more convenient to use the integrated rate
\begin{equation}
I'(h_a^x,h_b^y, \Delta t) = \int_0^\infty I(h_a^x,t_a; h_b^y, t_a+\Delta t) dt_a,
\end{equation}
which is simply given by $I(l^x_a, t_a; l^y_b, t_a+\Delta t)$ as in (\ref{il})  with $e^{-(\Gamma_S+\Gamma_L)t_a}$ replaced as $1/(\Gamma_S+\Gamma_L)$.

We focus on the situation that  $|h_a^+\rangle = |h_b^+\rangle=|h^+\rangle$ and  $|h_a^-\rangle = |h_b^-\rangle=|h^-\rangle$,  hence
\begin{eqnarray}
w_{+a}^+=w_{+b}^+  & \equiv  & Q^+ , \\
w_{-a}^{+}=w_{-b}^{+} &  \equiv& X^{+}, \\
w_{+a}^-=w_{+b}^-  & \equiv  & X^-, \\
w_{-a}^{-}=w_{-b}^{-} &  \equiv& Q^{-}.
\end{eqnarray}
It is straightforward to find the following properties. If CP is conserved directly, then $X^{\pm}=0$, hence  $X^{\pm}$ is a parameter  characterizing the direct CP violation.   If CPT is conserved directly, then $X^{\pm}$ is purely imaginary, i.e. $X^{\pm}=-{X^{\pm}}^*$.

We define
\begin{equation}
\xi_{\pm} \equiv \frac{X^{\pm}}{Q^{\pm}}.
\end{equation}

$Q^{\pm}$ and $X^{\pm}$ are not  directly measurable quantities, as $M_S$ and $M_L$, rather than $M_+$ and $M_-$, are physical. However, from (\ref{ms1}) and (\ref{ml1}), we have
\begin{eqnarray}
Q^++\epsilon_S X^+ &=&\sqrt{1+|\epsilon_S|^2} \langle h^+|{\cal H}|M_S\rangle, \\
\epsilon_L X^- + Q^-&=&\sqrt{1+|\epsilon_L|^2} \langle h^-|{\cal H}| M_L\rangle.
\end{eqnarray}
Therefore,
\begin{eqnarray}
\eta_{h^+} & \equiv &\frac{\langle h^+|{\cal H}|M_L\rangle}{\langle h^+|{\cal H}|M_S\rangle} = \frac{\xi^++\epsilon_L}{1+\epsilon_S \xi^+ },\\
\eta_{h^-} & \equiv & \frac{\langle h^-|{\cal H}|M_S\rangle}{\langle h^-|{\cal H}|M_L\rangle} = \frac{\xi^- +\epsilon_S}{1+\epsilon_L \xi^-}.
\end{eqnarray}
In the case of $|h^+\rangle = |\pi^+\pi^-\rangle$, $|\eta_{h^+}\rangle$ is just the well-known  $$\eta_{+-} = \frac{\langle \pi^+\pi^-|{\cal H}|M_L\rangle}{\langle \pi^+\pi^-|{\cal H}|M_S\rangle}.$$  In the case of   $|h^+\rangle = |\pi^0\pi^0\rangle$,  $|\eta_{h^+}\rangle$ is just the well-known  $$\eta_{00} = \frac{\langle \pi^0\pi^0|{\cal H}|M_L\rangle}{\langle \pi^0\pi^0|{\cal H}|M_S\rangle}.$$

One obtains
\begin{eqnarray}
M(h^+,h^+)  & = & -N(h^+,h^+)= \frac{1}{2\sqrt{2}(1+\Omega)}
[-(1 -x_L^{-1}+ x_S-\Omega)(Q^+)^2  + 2(x_L^{-1} + x_S) Q^+X^+
\nonumber \\&&  + (1 +x_L^{-1} - x_S-\Omega)(X^+)^2], \\
M(h^-,h^-)  & = & -N(h^-,h^-)= \frac{1}{2\sqrt{2}(1+\Omega)}
[-(1 -x_L^{-1}+ x_S-\Omega)(X^-)^2  + 2(x_L^{-1} + x_S) X^-Q^-
\nonumber \\&&  + (1 +x_L^{-1} - x_S-\Omega)(Q^-)^2], \\
M(h^+,h^-)  & = & -M(h^-,h^+)\nonumber \\
&=&  \frac{1}{2\sqrt{2}(1+\Omega)}[(1 + x_S) Q^+  + (1- x_S) X^+][(1 +x_L^{-1})Q^- - (1 - x_L^{-1})X^-] \equiv Z, \\
N(h^+,h^-)  & = & -N(h^-,h^+)\nonumber \\
 &= & \frac{1}{2\sqrt{2}(1+\Omega)}[(1 - x_L^{-1}) Q^+  - (1 +x_L^{-1})X^+][ (1- x_S)Q^- +(1 + x_S) X^-] \equiv Y.
\end{eqnarray}

Therefore, we obtain
\begin{equation}
\begin{array}{rcl}
I(h^+, t_a; h^+, t_b) & = &
\frac{| -(1 -x_L^{-1}+ x_S-\Omega)(Q^+)^2  + 2(x_L^{-1} + x_S) Q^+X^+
+ (1 +x_L^{-1} - x_S-\Omega)(X^+)^2   |^2  }{8|1+\Omega|^2}   \\ &&\times e^{-(\Gamma_S+\Gamma_L)t_a} [e^{-\Gamma_S \Delta t} + e^{-\Gamma_L\Delta t} - 2 e^{-\frac{1}{2}(\Gamma_S+\Gamma_L)\Delta t} \cos(\Delta m \Delta t)], \\
I(h^-, t_a; h^-, t_b) & = &
\frac{|-(1 -x_L^{-1}+ x_S-\Omega)(X^-)^2  + 2(x_L^{-1} + x_S) X^-Q^-
+ (1 +x_L^{-1} - x_S-\Omega)(Q^-)^2  |^2  }{8|1+\Omega|^2}   \\ &&\times e^{-(\Gamma_S+\Gamma_L)t_a} [e^{-\Gamma_S \Delta t} + e^{-\Gamma_L\Delta t} - 2 e^{-\frac{1}{2}(\Gamma_S+\Gamma_L)\Delta t} \cos(\Delta m \Delta t)], \\
I(h^+, t_a; h^-, t_b) & = &  e^{-(\Gamma_S+\Gamma_L)t_a}
\{|Z|^2 e^{-\Gamma_L\Delta t}+ |Y|^2 e^{-\Gamma_S\Delta t} + 2e^{-\frac{\Gamma_S+\Gamma_L}{2}\Delta t} [\Re (Z^*Y)\cos (\Delta m \Delta t) + \Im (Z^*Y) \sin  (\Delta m \Delta t) ]\},
\\
I(h^-, t_a; h^+, t_b) &=&   e^{-(\Gamma_S+\Gamma_L)t_a}
\{|Y|^2 e^{-\Gamma_L\Delta t}+ |Z|^2 e^{-\Gamma_S\Delta t} + 2e^{-\frac{\Gamma_S+\Gamma_L}{2}\Delta t} [\Re (Z^*Y)\cos (\Delta m \Delta t) - \Im (Z^*Y) \sin  (\Delta m \Delta t) ]\}.
\end{array} \label{i2}
\end{equation}

One can also consider the integrated rate
\begin{equation}
I'(h^x, h^y, \Delta t) = \int_0^\infty I(h^x, t_a; h^y, t_a+\Delta t) dt_a, \label{i22}
\end{equation}
which is simply given by $I(h^x, t_a; h^y, t_a+\Delta t)$ as in (\ref{i2}),  with $e^{-(\Gamma_S+\Gamma_L)t_a}$ replaced as $1/(\Gamma_S+\Gamma_L)$.

\section{CP and CPT violations in  decays in CP basis   \label{cp2}}

First,  it can be seen that the equal-CP decay rates $I[h^+, t_a; h^+, t_b]$ and $I[h^-, t_a; h^-, t_b]$ are both proportional
$  e^{-\Gamma_S \Delta t} + e^{-\Gamma_L \Delta t} - 2 e^{-\frac{1}{2}(\Gamma_S+\Gamma_L)\Delta t} \cos(\Delta m \Delta t),$ as demonstrated by KLOE experimental data~\cite{domenico,kloe}.

For $\Delta t=0$, we always have $I[h^+, t_a; h^+, t_a]=[h^-, t_a; h^-, t_a]=I'[h^+, h^+, 0]=[h^-,  h^-,0]=0$   no matter whether CP or CPT is violated.

The four joint rates in (\ref{i2}) or (\ref{i22}) can form some asymmetries between different modes of decays into CP eigenstates. As
$
I'(h^x, h^y, \Delta t)e^{-(\Gamma_L+\Gamma_S)t_a} = \frac{1}{\Gamma_L+\Gamma_S} I(h^x, t_a; h^y, t_b),$
the asymmetries defined for the instantaneous joint probability $I$
and its integration $I'$ for a specific $\Delta t$ are equal, and depend  only  on $\Delta t \equiv t_b-t_a$,
\begin{equation}
B(h^xh^y, h^yh^x, \Delta t) \equiv \frac{I[h^x,t_a; h^y,t_a+\Delta t ]-I[h^y,t_a; h^x,t_a+\Delta t]}{I[h^x,t_a; h^y,t_a+\Delta t]+ I[h^y,t_a; h^x,t_a+\Delta t]}=  \frac{I'[h^x, h^y, \Delta t]-I'[h^y, h^x, \Delta t]}
{I'[h^x, h^y, \Delta t]+I'[h^y, h^x, \Delta t]},
\end{equation}
which is always time-independent.

In general, the equal-CP asymmetry  is found to be
\begin{equation}
B(++,--, \Delta t) = \frac{ P(Q^+,X^+)  - P(X^-,Q^-)  }{P(Q^+,X^+) + P(X^-,Q^-)  }, \label{bb10}
\end{equation}
where
\begin{equation}
P(\beta,\gamma) \equiv | -(1 -x_L^{-1}+ x_S-\Omega)\beta^2  + 2(x_L^{-1} + x_S)\beta\gamma + (1 +x_L^{-1} - x_S-\Omega)\gamma^2|^2.
\end{equation}

{\bf Theorem 5} {\em The equal-CP asymmetry  $B(++,--, \Delta t)$ is always a constant independent of $\Delta t$. }

In general, the unequal-CP asymmetry
is found to be
\begin{eqnarray}
B(+-,-+, \Delta t)& =& \frac{(|Z|^2-|Y|^2) ( e^{-\Gamma_L\Delta t} -e^{-\Gamma_S\Delta t})+ 4 e^{-\frac{\Gamma_S+\Gamma_L}{2}\Delta t} \Im(Z^*Y) \sin (\Delta m \Delta t) }{(|Z|^2+|Y|^2) ( e^{-\Gamma_L\Delta t} +e^{-\Gamma_S\Delta t})+ 4 e^{-\frac{\Gamma_S+\Gamma_L}{2}\Delta t} \Re(Z^*Y) \cos (\Delta m \Delta t) } \\
& =& \frac{(|\tilde{Z}|^2-|\tilde{Y}|^2) ( e^{-\Gamma_L\Delta t} -e^{-\Gamma_S\Delta t})+ 4 e^{-\frac{\Gamma_S+\Gamma_L}{2}\Delta t} \Im(\tilde{Z}^*\tilde{Y}) \sin (\Delta m \Delta t) }{(|\tilde{Z}|^2+|\tilde{Y}|^2) ( e^{-\Gamma_L\Delta t} +e^{-\Gamma_S\Delta t})+ 4 e^{-\frac{\Gamma_S+\Gamma_L}{2}\Delta t} \Re(\tilde{Z}^*\tilde{Y}) \cos (\Delta m \Delta t) }, \label{une}
\end{eqnarray}
where
\begin{eqnarray}
\tilde{Z} & \equiv & [(1 + x_S)  + (1- x_S) \xi^+][(1 +x_L^{-1})- (1 - x_L^{-1})\xi^-], \\
\tilde{Y} & \equiv & [(1 - x_L^{-1})    - (1 +x_L^{-1}) \xi^+][ (1- x_S)+(1 + x_S) \xi^-]. \label{zy1}
\end{eqnarray}
When $\Delta t=0$, the unequal-CP asymmetry is always $0$    no matter whether CP or CPT is violated. Therefore we have the following result.

{\bf Theorem 6} {\em For $\Delta t =0$, the unequal-CP asymmetry $B(+-,-+, 0)$ vanishes,   no matter whether CP or CPT is violated.  }

If CP is conserved directly, then $X^{\pm}=0$,  consequently  $B(++,--, \Delta t)$ reduces to
\begin{eqnarray}
B(++,--,\Delta t)
=  \frac{{|Q^+|}^4|(1 - x_L^{-1})(1 + x_S)|^2 - {|Q^-|}^4 |(1 +x_L^{-1})(1 - x_S)|^2 }{ {|Q^+|}^4|(1 - x_L^{-1})(1 + x_S)|^2 + {|Q^-|}^4 |(1 +x_L^{-1})(1 - x_S)|^2   }. \label{eqdirect} \end{eqnarray}
while $B(+-,-+,  \Delta t)$ reduces to
\begin{eqnarray}
B(+-,-+,  \Delta t) = \frac{(1-|W|^2) ( e^{-\Gamma_L\Delta t} -e^{-\Gamma_S\Delta t})+ 4 e^{-\frac{\Gamma_S+\Gamma_L}{2}\Delta t}\Im(W) \sin (\Delta m \Delta t) }{(1+|W|^2) ( e^{-\Gamma_L\Delta t} +e^{-\Gamma_S\Delta t})+ 4 e^{-\frac{\Gamma_S+\Gamma_L}{2}\Delta t}  \Re(W)\cos (\Delta m \Delta t) }, \label{nod}
\end{eqnarray}
where $W\equiv  \frac{(1-x_L^{-1})(1-x_S)}{(1+x_L^{-1})(1+x_S)}$.

If CP is conserved indirectly, no matter whether CPT is conserved indirectly, we have
$x_S=x_L^{-1} =1$.   Consequently,
$M(h^+,h^+)   =  -N(h^+,h^+)= \frac{1}{\sqrt{2}}  Q^+X^+,$
$M(h^-,h^-)   =  -N(h^-,h^-)= \frac{1}{\sqrt{2}} X^-Q^-,$
$Z= M(h^+,h^-)   =  -M(h^-,h^+)=  \frac{1}{\sqrt{2}} Q^+ Q^- $,
$Y= N(h^+,h^-)   =  -N(h^-,h^+) = -\frac{1}{\sqrt{2}} X^+  X^-$.
Therefore
\begin{equation}
\begin{array}{rcl}
I(h^+, t_a; h^+, t_b) & = &  \frac{|Q^+X^+|^2}{2}e^{-(\Gamma_S+\Gamma_L)t_a} [e^{-\Gamma_S \Delta t} + e^{-\Gamma_L\Delta t} - 2 e^{-\frac{1}{2}(\Gamma_S+\Gamma_L)\Delta t} \cos(\Delta m \Delta t)], \\
I(h^-, t_a; h^-, t_b) & = &\frac{|Q^-X^-|^2}{2}e^{-(\Gamma_S+\Gamma_L)t_a} [e^{-\Gamma_S \Delta t} + e^{-\Gamma_L\Delta t} - 2 e^{-\frac{1}{2}(\Gamma_S+\Gamma_L)\Delta t} \cos(\Delta m \Delta t)], \\
I(h^+, t_a; h^-, t_b) & = & \frac{1}{2}  e^{-(\Gamma_S+\Gamma_L)t_a}
\{|Q^+Q^-|^2 e^{-\Gamma_L\Delta t}+ |X^+X^-|^2 e^{-\Gamma_S\Delta t} \\
&&+ 2e^{-\frac{\Gamma_S+\Gamma_L}{2}\Delta t} [\Re ((Q^+Q^-)^*X^+X^-)\cos (\Delta m \Delta t) + \Im ((Q^+Q^-)^*X^+X^-) \sin  (\Delta m \Delta t) ]\},
\\
I(h^-, t_a; h^+, t_b) &=& \frac{1}{2}  e^{-(\Gamma_S+\Gamma_L)t_a}
\{|X^+X^-|^2 e^{-\Gamma_L\Delta t}+ |Q^+Q^-|^2 e^{-\Gamma_S\Delta t}  \\&&+ 2e^{-\frac{\Gamma_S+\Gamma_L}{2}\Delta t} [\Re ((Q^+Q^-)^*X^+X^-)\cos (\Delta m \Delta t) - \Im ((Q^+Q^-)^*X^+X^-) \sin  (\Delta m \Delta t) ]\},
\end{array} \label{indirect2}
\end{equation}

Therefore if CP is conserved indirectly, no matter whether CPT is conserved indirectly,  the equal-CP asymmetry is simplified to
\begin{equation}
B(++,--, \Delta t) = \frac{|Q^+X^+|^2- |Q^-X^-|^2  }{|Q^+X^+|^2+ |Q^-X^-|^2  },  \label{eqcpas}
\end{equation}
while the unequal-CP asymmetry is simplified to
\begin{eqnarray}
B(+-,-+, \Delta t)& =& \frac{(1-|\xi^+\xi^-|^2) ( e^{-\Gamma_L\Delta t} -e^{-\Gamma_S\Delta t})+ 4 e^{-\frac{\Gamma_S+\Gamma_L}{2}\Delta t} \Im(\xi^+\xi^-) \sin (\Delta m \Delta t) }{( 1+|\xi^+\xi^-|^2  ) ( e^{-\Gamma_L\Delta t} +e^{-\Gamma_S\Delta t})+ 4 e^{-\frac{\Gamma_S+\Gamma_L}{2}\Delta t} \Re(\xi^+\xi^- ) \cos (\Delta m \Delta t) }. \label{uneq2}
\end{eqnarray}

If CP is conserved both directly and indirectly,  no matter whether CPT is conserved or not, we have $X^{\pm}=0$, $x_S=x_L^{-1}=1$, then
\begin{equation}
\begin{array}{rcl}
I(h^+_a, t_a; h^+_b, t_b) & = &0, \\
I(h^-_a, t_a; h^-_b, t_b) & = &0, \\
I(h^+_a, t_a; h^-_b, t_b) & = & \frac{|Q^+Q^-|^2}{2} e^{-(\Gamma_S+\Gamma_L)t_a} e^{-\Gamma_L\Delta t},
\\
I(h^-, t_a; h^+, t_b) &=& \frac{|Q^+Q^-|^2}{2}   e^{-(\Gamma_S+\Gamma_L)t_a}  e^{-\Gamma_S\Delta t},
\end{array} \label{i2cp}
\end{equation}
As  $I(h^+_a, t_a; h^+_b, t_b)=I(h^-_a, t_a; h^-_b, t_b)=I'(h^+_a, h^+_b, \Delta t)=I(h^-_a,  h^-_b,\Delta t)=0$, it is meaningless to define equal-CP asymmetry in this case.  In this case, the unequal-CP asymmetry is
\begin{eqnarray}
B(+-,-+,  \Delta t) = \frac{ e^{-\Gamma_L\Delta t} -e^{-\Gamma_S\Delta t} }{  e^{-\Gamma_L\Delta t} +e^{-\Gamma_S\Delta t}  }, \label{noid}
\end{eqnarray}

{\bf Theorem 7} {\em If any equal-CP joint decay rate is nonzero, then CP must be violated,  directly or indirectly or both.   }

If CPT is conserved indirectly,  then   $ \delta_M=0$, thus $x_S=x_L =q/p$, and thus  $\Omega =1$, consequently  the equal-CP asymmetry is given by (\ref{bb10}) with $P(\beta,\gamma)$ simplified to
\begin{equation}
P(\beta,\gamma) \equiv |2(\frac{p}{q} + \frac{q}{p})\beta\gamma + (\frac{p}{q} -\frac{q}{p})(\gamma^2-\beta^2)|^2. \label{eqcpt}
\end{equation}
The unequal-CP asymmetry
$B(+-,-+,  \Delta t)$ is given by (\ref{une}) with $\tilde{Z}$ and $\tilde{Y}$ reduced to
\begin{eqnarray}
\tilde{Z} & \equiv &   [(1 + \frac{q}{p})  + (1- \frac{q}{p}) \xi^+][(1 +\frac{p}{q} )- (1 - \frac{p}{q} )\xi^-], \\
\tilde{Y} & \equiv & [(1 - \frac{p}{q} )    - (1 +\frac{p}{q} ) \xi^+][ (1- \frac{q}{p})+(1 + \frac{q}{p}) \xi^-]. \label{zy2}
\end{eqnarray}

If   CPT is  conserved  indirectly while CP is conserved directly, $x_S=x_L=q/p$ while  $X^{\pm}=0$, then the equal-CP asymmetry  reduces to
\begin{eqnarray}
B(++,--, \Delta t)
 =  \frac{{|Q^+|}^4 - {|Q^-|}^4 }{ {|Q^+|}^4 + {|Q^-|}^4   },  \label{bab}
 \end{eqnarray}
while  the unequal-CP asymmetry
$B(+-,-+,  \Delta t)$ is given by (\ref{nod}) now with $W =  \frac{2-(\frac{p}{q}+\frac{q}{p})}{2+(\frac{p}{q}+\frac{q}{p})}$.

If CPT is conserved directly, then  $X^{\pm}$ is purely imaginary, all the above results under other additional conditions are still valid, respectively,  under  the constraint that  $X^{\pm}$ is purely imaginary.

\section{summary \label{summary} }

In this paper, we have studied the joint decays of a pair of two pseudoscalar neutral mesons in an entangled state of $C=-1$, as  produced in $\phi$ and $B$ factories. We exactly  calculated   the rates of the joint  decays   into flavor eigenstates, taking into account direct CP violation and violation of $\Delta {\cal F} =\Delta Q$ rule.   We obtained some exact results on how to extract information on CP and CPT violations from various asymmetries of the joint decays to flavor eigenstates, or joint decays to CP eigenstates.  Measurement of such joint rates and asymmetries can be used to determine various parameters, including those of CP and CPT violations.

Remarkably, the special property of the entanglement of the $C=-1$ pair allows us to propose the  meaningful and useful definition of the transition amplitude between a  CP eigenstate of the meson and the decay product which is a CP eigenstate.

The equal-flavor asymmetry $A(++,--,\Delta t)$ and the equal-CP asymmetry $B(++,--,\Delta t)$ are both  always independent of $\Delta t$,  while the unequal-flavor asymmetry $A(+-,-+,  \Delta t)$ and the unequal-CP asymmetry $B(+-,-+,\Delta t)$ generically depend on  $\Delta t$.

We have considered various cases of  possible direct or indirect CP or CPT  violation, and obtain exact expressions of various asymmetries in various cases. Some of these exact results lead to simple yet powerful conclusions stated as several theorems.

If the equal-flavor asymmetry $A(++,--,\Delta t) \neq 0$,  then we have one or two of the following violations: (1) CP is violated indirectly, (2) both CP and CPT are violated directly. Therefore if one assumes direct $CPT$ conservation,  $A(++,--, \Delta t)\neq 0$ implies that CP must be violated indirectly.

Moreover, if   $A(++,--,\Delta t) \neq 0$ while CPT is assumed to be conserved both directly and indirectly, then in addition to indirect CP violation, we can draw the following conclusions: (1)   T must also be  violated indirectly; (2)   $|\langle l^+ |{\cal H}|\bar{M}^0\rangle| = |\langle  l^-|{\cal H}  |M^0\rangle \neq |\langle l^+|{\cal H}  |M^0\rangle| = |\langle l^- |{\cal H}|\bar{M}^0\rangle| $.

If  the unequal-flavor asymmetry
$A(+-,-+,  \Delta t) \neq 0 $, then CP must be violated  directly or indirectly.

Moreover, if   $A(+-,-+,  \Delta t)\neq 0$ for $\Delta t \neq 0$ while CPT is assumed to be conserved both directly and indirectly, then  we can draw the following conclusions: (1)     $|\langle l^+ |{\cal H}|\bar{M}^0\rangle| = |\langle  l^-|{\cal H}  |M^0\rangle \neq |\langle l^+|{\cal H}  |M^0\rangle| = |\langle l^- |{\cal H}|\bar{M}^0\rangle| $; (2) $\langle l^{-}|{\cal H} |M^0\rangle = \langle l^{+}|{\cal H}|\bar{M}^0\rangle^* \neq 0$, which means $\Delta {\cal F} =\Delta Q$ rule must be violated.

For joint decays to CP eigenstates, in addition to various detailed results of the asymmetries,  we have the following conclusions. The equal-time equal-CP rates $I[h^+,t_a;h^+,t_a]$ and $I[h^-,t_a;h^-,t_a]$  vanish, no matter whether CP or CPT is violated. At equal times $\Delta t =0$, the unequal-CP asymmetry $B(+-,-+, 0)$ vanishes,   no matter whether CP or CPT is violated. On the other hand, if any equal-CP decay rate is nonzero, then CP must be violated.

The detailed expressions of the rates and symmetries can be used to determine the CP and CPT violating parameters.   Clearly, these results are consequences of the well-known antisymmetry and entanglement of the state $|\Psi_-\rangle$. We hope these exact results are useful in studies on CP and CPT violations.

\vspace{1cm}

I am grateful to A. Di Domenico and other members of KLOE2, as well as Z. Huang, for useful discussions.
This work was supported by the National Science Foundation of China (Grant No. 10875028).

\end{document}